\newcommand{\beq}{\begin{equation}}
\newcommand{\enq}{\end{equation}}
\newcommand{\bee}{\begin{eqnarray}}
\newcommand{\ene}{\end{eqnarray}}
\newcommand{\non}{\nonumber}
\newcommand{\bq}{{\bf q}}
\newcommand{\bl}{{\bf l}}

\newcommand{\bn}{{\bf n}}
\newcommand{\cn}{{\mathcal N}}
\newcommand{\tcn}{\tilde{{\mathcal N}}}

\documentclass{article}

\begin{document}
\textwidth 16cm

\title{RESONANT MODE INTERACTIONS  \\ IN NONLINEAR
LATTICES}
 
\author{V.V. Konotop \\ 
Departamento de F\'{\i}sica and CFMC,  \\  Universidade de Lisboa, Complexo
Interdisciplinar, \\ Av. Prof. Gama Pinto 2,  
 Lisbon 1649-003, \\ Portugal}

%\email{konotop@cii.fc.ul.pt}

\maketitle

\begin{abstract}
Resonant mode interactions in weakly nonlinear multi-dimensional
lattices and related effects are described. We concentrate on
formal description of the phenomenon and consider as examples
mode interactions and  evolution equations for quadratic solitons
in two-dimensional lattices, and coherent structures in a
double-chain as examples of the application of the theory.

\end{abstract}

\section{Introduction}

Elastic interactions among quasi-particles, like phonons,
electrons, magnons, etc.,  in solids constitute fundamental
phenomena allowing one to observe microscopic effects on the
macroscopic level. Today they are very well studied and are
described in numerous text-books (see e.g. Modelung, 1981).
Although interactions mean nonlinearity and thus possibility of a
complex dynamics such phenomena have been discussed, so far, on
the language of quasi-particles. This means that usually only
two-body interactions are taken into account. Meantime, due to
fundamental discoveries of sixties and early seventies, namely
due to construction of the modern soliton theory, it became clear
that the dynamic of nonlinear systems is much richer than one
provided by the theory of two-body interactions. Namely, it
became clear that together with linear quasi-particles one should
consider solitons (or solitary waves) which are essentially
nonlinear objects and cannot be constructed from the linear
theory by any perturbation technique. In particular, speaking
about nonlinear lattices one should consider not only phonons and
interactions among them, but also solitary waves, called envelope
solitons (Tsurui, 1972). Meantime solitons are approximate
solutions of the nonlinear lattices and they are necessarily
accompanied by generation of higher harmonics (Jimenez, 1999).
The fact that corresponding dynamics displays indeed a complicate
behavior has been reported in (Economou, {\it at al}  1994;
Kopidakis, {\it at al.} 1994). The effect of high harmonics
becomes especially important when the resonant conditions
\begin{equation}
\label{threewave} \omega_3=\omega_1+\omega_2, \qquad
\bq_3=\bq_1+\bq_2+{\bf Q}
\end{equation}
or
\begin{equation}
\label{second} \omega_2=2\omega_1, \qquad \bq_2=2\bq_1+{\bf Q}
\end{equation}
are satisfied. On the language of solid state physics one can
interpret the first and second equalities in (\ref{threewave}) or
in (\ref{second}) as conservation laws of energy and momentum for
phonons, respectively. Using the terminology of the nonlinear
optics these formulas can be called matching conditions or
conditions of resonant three mode generation and resonant
generation of the second harmonic. The present work describes
various phenomena in nonlinear multi-dimensional lattices when
one of the above resonance conditions is satisfied.

The theory we develop is applied to lattices with a complex cell,
to vector lattices (with a simple or complex cell), or to couple
lattices of different physical nature (like for example to a
system describing electron-phonon interactions (Konotop, 1997)).
Also, subject to certain modification the theory can be adjusted
to describe coupled nonlinear chains. In the context of the
present work the last system is interesting since it does not
belong either class of one-dimensional nor two-dimensional
lattices. Also a double chain is an example of a system when one
can provide resonant conditions between localized and extended
degrees of freedom.

\section{Properties of a linear lattice.}

Let us recover some general properties of a linear $\cn$D lattice
[hereafter $\cn$ ($=1,2,3$) indicates the dimension of the
space]. It will be assumed that the lattice is finite with $N_j$
($j=1,...,\cn$) cells in $j$th direction, so that the total number
of cells is $N = N_1\cdots N_\cn$. It also will be assumed that
each cell has $\alpha$ generalized degrees of freedom: they can be
related either to different atoms, or to different components of
the displacement vector. After all we will be interested in the
limit $N_j \to \infty$. We start with a generic case, when the
evolution equation is written as follows
\beq \label{evolut} \ddot{\Psi}({\bf n}) = - \sum_{{\bf l}}J({\bf
n},{\bf l})\Psi({\bf l}). \enq
Here $ {\Psi}({\bf n}) $ is a $\tcn \times \tcn$ matrix with
$\tcn=\alpha\cn$ having elements ${\psi}_{ij}({\bf n})$, ${\bf
n}$ is an $\cn$-dimensional vector indicating the position of a
cell: it can be represented in the standard form ${\bf n}=\sum_j
n_j{\bf e}_j$, where ${\bf e}_j$ are lattice vectors with $n_j=
0,1\ldots N_j-1$,  $ J({\bf n},{\bf l})$ is a $\tcn \times \tcn$
matrix of force constants which is considered to be hermitian, $
J_{ij}({\bf n},{\bf l}) = \bar{J}_{ji}({\bf n},{\bf l})$ (the
overbar stands for the complex conjugation) and having only
positive eigenvalues, and a dot standing for the derivative with
respect to time. It will be assumed that the second index in
$\psi_{ij}$ refers to a branch of the spectrum, while the first
one is associated with a generalized degree of freedom. Lattice
(\ref{evolut}) is considered subject to periodic boundary
conditions $\Psi({\bf n} + N_{j}{\bf e}_{j}) = \Psi({\bf n})$ for
all $j$.

Evolution equation (\ref{evolut}) can be associated with the
spectral problem
\beq \label{spectral3} \Phi({\bf n};{\bf q})\Lambda({\bf q}) =
\sum_{{\bf l}}J({\bf n},{\bf l})\Phi({\bf l};{\bf q}) \enq
where ${\Phi}({\bf n};{\bf q}) = (\Phi_1({\bf n};{\bf q}) ,\ldots,
\Phi_{\tcn}({\bf n};{\bf q}))$: $\Phi_{\eta}({\bf n};{\bf q})$ are
column eigenfunctions, $
 \Lambda({\bf q}) =
\mbox{diag}[\lambda_{1}({\bf q}),\ldots ,\lambda_{\tcn}({\bf
q})]\,, $ is the eigenvalue matrix: $\omega^{2}_{\eta}({\bf q})$
$\equiv\lambda_{\eta}({\bf q})$, (${\eta} = 1\ldots \tcn$) and
${\bf q}$ is a vector in the reciprocal space. It will be the
assumption that in the spectrum of the linear eigenvalue problem
there are no degenerate points i.e. $\lambda_\eta({\bf q}) \neq
\lambda_{\eta'}({\bf q})$ for any ${\bf q}$ if $\eta \neq \eta'$.

We will be interested in translationally invariant lattices, this
is
%\beq \label{symmetry1}
$J({\bf n},{\bf l})$ $=$ $J(\mid{\bf n} - {\bf l}\mid). $
%\enq
This property allows one to represent the eigenfunctions in the
form
\beq \label{phi} \Phi({\bf n},{\bf q}) = A({\bf q})e^{i{\bf
q}{\bf n}} \enq
where $A({\bf q})$ is a matrix depending on the wave vector only.
It solves the equation
\beq \label{spectral2}  A({\bf q})\Lambda({\bf q}) = J({\bf
q})A({\bf q}) \enq
with
%\beq \label{fourier_J}
$
J({\bf q}) = \sum_{{\bf n}}J(\mid{\bf n}\mid)e^{-i{\bf q}{\bf
n}}.
$
%\enq
%
  It will be
assumed that the eigenfunctions $\phi_{\eta}({\bf n},{\bf q})$
make up an orthonormal complete set. Thus
\beq
%\label{normal}
 \sum_{{\bf n}}\Phi^{\dagger}({\bf n};{\bf
q}')\Phi({\bf n};{\bf q}) = \Delta({\bf q},{\bf q}'),
%\enq
%
%and
%
%\beq
\qquad
  \sum_{{\bf q}}\Phi({\bf n};{\bf q})\Phi^{\dagger}({\bf
l};{\bf q}) = \Delta({\bf n},{\bf l}) \label{closure} \enq
where $\Delta ({\bf q},{\bf q}')= I\delta_{{\bf q},{\bf q}'}$,
$\delta_{{\bf q},{\bf q}'}$ is the Kronecker delta, $I$ is a unit
$4\times 4$ matrix and the dagger stands for the Hermitian
conjugate. It follows from (\ref{closure}) that $A({\bf q})$ is a
unitary matrix
%\beq
$
A^{\dagger}({\bf q})A({\bf q}) = I
$
%\enq
%
which means that $\Phi({\bf n},{\bf q}) $ is a unitary matrix, as
well. Now the dispersion relation of the linear lattice can be
expressed as
\beq \label{expression} \Lambda({\bf q}) = \sum_{{\bf
n}}\sum_{{\bf l}}\Phi^{\dagger}({\bf n};{\bf q})J({\bf n},{\bf
l})\Phi({\bf l};{\bf q}). \enq

Next we express the group velocity through the eigenfunctions of
the linear spectral problem. To this end we consider a small
variation ${\bf q} \longrightarrow {\bf q} + d{\bf q}$ and the
resulting changes of the eigenfunctions
\[ \Phi({\bf l};{\bf q}) \longrightarrow \Phi({\bf l};{\bf q}) +
\left[i{\bf l}\cdot d{\bf q} - \frac 12 {({\bf l}\cdot d{\bf
q})^2}\right]\Phi({\bf l};{\bf q})  \]
This variation can be interpreted as one caused by the
perturbation
\beq \label{perturb} \delta J({\bf n},{\bf l}) = \left[i({\bf l} -
{\bf n})\cdot d{\bf q} - {1\over2}(({\bf l} - {\bf n})\cdot d{\bf
q})^2\right]J({\bf n},{\bf l}) \enq
of the Kernel $J({\bf n},{\bf l})$. Thus on the one hand the
Taylor expansion yields
\beq \label{taylor} \Lambda({\bf q} + d{\bf q}) = \Lambda({\bf q})
+ (d{\bf q}\nabla_q) \Lambda ({\bf q}) + \frac 12 (d{\bf
q}\nabla_q)^2 \Lambda ({\bf q})+ O(|d{\bf q}|^3) \enq
where $\nabla_q\equiv\partial /\partial {\bf q}$. On the other
hand we obtain from the standard perturbation technique
\[ \Lambda({\bf q} + d{\bf q}) = \Lambda({\bf q})
+ \Lambda^{(1)}({\bf q}) + \Lambda^{(2)}({\bf q})+ O(|d{\bf q}|^3)
\]
where the diagonal matrices $\Lambda^{(j)}({\bf q})$ have
elements as follows
\beq \label{lambda1} \Lambda^{(1)}_{\eta\eta}({\bf q}) =
i\sum_{{\bf l},{\bf n}} (d{\bq}\cdot({\bf n}-{\bf l}))
K_{\eta\eta}(\bn,\bl;{\bf q}) \enq
\bee \label{lambda2} \Lambda^{(2)}_{\eta\eta}({\bf q}) =
\sum_{{\bf l},{\bf n}} (d{\bq}\cdot({\bf n}-{\bf
l}))^2K_{\eta\eta}(\bn,\bl;\bq)  \non \\
-\sum_{{\bf l},{\bf n}}\sum_{{\bf l}',{\bf n}'} (d\bq({\bf
n}-{\bf l})) (d\bq ({\bf n}'-{\bf l}')) \sum_{\alpha\neq\eta}
\frac{ K_{\eta\alpha}(\bn,\bl;{\bf
q})K_{\alpha\eta}(\bn',\bl';{\bf
q})}{\lambda_\eta-\lambda_\alpha} \ene
where we have introduced the matrix
 \beq \label{k}  K_{\eta\alpha}(\bn,\bl;{\bf q}) =
 \Phi^{\dagger}_{\eta}({\bf n};{\bf q})
J({\bf n},{\bf l}) \Phi_{\alpha}({\bf l};{\bf q})  \enq

Comparing (\ref{lambda1}) and (\ref{lambda2}) with (\ref{perturb})
we compute the group velocity
\beq \label{gv} {\bf v}_\eta({\bf q}) \equiv  \nabla_q
\omega_{\eta}({\bf q})= \frac{i} {2\omega_\eta({\bf q})}
\sum_{{\bf n_1},{\bf n}_2}({\bf n}_2 - {\bf n}_1)K_{\eta\eta}({\bf
n}_1,{\bf n}_2;{\bf q}). \enq
and the tensor of the group velocity dispersion
\bee \label{tgvd}\omega_{\eta}^{ij}
%\equiv \frac{\partial^2\omega_\eta}{\partial q_i\partial q_j}
=
-\frac{1}{\omega_\eta}\frac{\partial\omega_\eta}{\partial q_i}
\frac{\partial\omega_\eta}{\partial q_j}+
\frac{1}{\omega_\eta}\sum_{{\bf l},{\bf n}}
(n_i-l_i)(n_j-l_j)K_{\eta\eta}(\bn,\bl;\bq)
\non \\
-\frac{1}{\omega_\eta}\sum_{{\bf l},{\bf n}}\sum_{{\bf l}',{\bf
n}'} (n_i-l_i) (n_j^{\prime}-l_j^{\prime}) \sum_{\alpha\neq\eta}
\frac{ K_{\eta\alpha}(\bn,\bl;{\bf
q})K_{\alpha\eta}(\bn',\bl';{\bf
q})}{\lambda_\eta-\lambda_\alpha} \ene

\section{Nonlinear model}

The subject of our main interest is the resonant mode interactions
in a nonlinear lattice possessing quadratic nonlinearity. As in
the previous section we start with the general model whose
dynamics is governed by the equation
\bee \label{nlevol} \ddot{\Psi}({\bf n}) = - \sum_{{\bf
n}_1}J({\bf n},{\bf n}_1) \Psi({\bf n}_1)
%\non\\
-\sum_{{\bf n}_1,{\bf n}_2}\hat{J}_3({\bf n},{\bf n}_1,{\bf n}_2):
\Psi({\bf n}_1)\Psi({\bf n}_2) \ene
Here $\Psi({\bf n})=$col$(\psi_1(\bn),\ldots,,\psi_{\tcn}(\bn))$,
$J({\bf n},{\bf n}_1)$ is a $\tcn \times \tcn$ matrix possessing
the properties described in the previous section and
$\hat{J}_3({\bf n},{\bf n}_1,{\bf n}_2)$ is understood to be a
tensor with components $J_{3,jj_1j_2}$, such that the last term in
the formal expression in (\ref{nlevol}) means that the $j$th
component of the nonlinear term is given by
\[ \left(\hat{J}_3({\bf n},{\bf n}_1,{\bf n}_2): \Psi({\bf
n}_1)\Psi({\bf n}_2)\right)_j =
%\\
\sum_{j_1,j_2}J_{3,jj_1j_2} ({\bf n},{\bf n}_1,{\bf
n}_2)\psi_{j_1}({\bf n}_1)\psi_{j_2}({\bf n}_2) \]
Below we concentrate on the case of nearest neighbor interactions.

In order to describe mode interactions we employ the multiscale
analysis. To this end we introduce a formal small parameter
$\epsilon$, $\epsilon\ll 1$, which generates a set of temporal
variables $t_{\nu} = \epsilon^{\nu}t ,\,\,\,\,\, (\nu =
0,1,\ldots)$ and a set of spatial variables ${\bf n}_0$ and ${\bf
r}_{\nu} = \epsilon^{\nu}{\bf n} ,\,\,\,\,\,(\nu=1,2,\ldots)$
which are regarded as independent. ${\bf r}_{\nu}$ will be
considered as a continuous variable. Respectively, we look for
the solution of (\ref{nlevol}) in the form of the expansion
\beq \label{expan} \Psi({\bf n}) =\epsilon \Psi^{(0)}({\bf n}_0,
\{{\bf r}_{\nu}\};\{t_{\nu}\}) + \epsilon^2 \Psi^{(1)}({\bf n}_0,
\{{\bf r}_{\nu}\};\{t_{\nu}\}) + \cdots \enq
where $\{{\bf r}_{\nu}\}$ and $\{t_{\nu}\}$ stands for the whole
set of variables. Being interested in the process of the
three-mode interactions we represent
\beq \label{e1} \psi^{(0)}({\bf n}_0, \{{\bf
r}_{\nu}\};\{t_{\nu}\}) =
%\non\\
\sum_{\alpha = 1}^3{\mathcal A}_{\alpha}({\bf r}_{1};t_{1})
\phi_{\eta_\alpha}({\bf n}_0;{\bf
q}_\alpha)e^{-i\omega_{\eta_\alpha}({\bf q}_\alpha)t_0} + c.c.
\enq
Here, $\phi_{\eta_\alpha}({\bf n}_0;{\bf q}_\alpha)$ are the
eigenfunctions of the eigenvalues $\omega^2_{\eta_\alpha}$ linked
by (\ref{threewave}), ${\mathcal A}_{\alpha}({\bf r}_{1};t_{1})$
can be interpreted as the amplitude of the envelope of the
$\alpha$-th mode and "c.c." stands for complex conjugation. In a
generic situation the amplitude ${\mathcal A}$ depends on all
"slow variables" $({\bf r}_1, {\bf r}_2, \ldots; t_1, t_2,
\ldots)$. However being interested in the process of three wave
interaction which corresponds to the scales defined by ${\bf r}_1$
and $t_1$ we indicate only these variables in the argument of the
amplitude. The scaling imposed implies neglecting the group
velocity dispersion.

Substituting the new variables in evolution equation
(\ref{nlevol}) and gathering terms of the same order of the
parameter $\epsilon$ one recovers the dispersion relation
(\ref{expression}) in the first order. In the second order with
respect to $\epsilon$ subject to the matching condition
(\ref{threewave}) one obtains the following system of equations
for slowly varying amplitudes
\beq \label{a1} \frac{\partial {\mathcal A}_1}{\partial t_1} +
({\bf v}_{\eta_1}\cdot\nabla_r){\mathcal A}_1 =
i\frac{\chi_{\eta_1\eta_2\eta_3}({\bf q}_1,{\bf
q}_2)}{\omega_{\eta_1}} \bar{{\mathcal A}}_2{\mathcal A}_3 \enq
\beq \label{a2} \frac{\partial {\mathcal A}_2}{\partial t_1} +
({\bf v}_{\eta_2}\cdot \nabla_r){\mathcal A}_2 =
i\frac{\chi_{\eta_1\eta_2\eta_3}({\bf q}_1,{\bf
q}_2)}{\omega_{\eta_2}} \bar{{\mathcal A}}_1{\mathcal A}_3 \enq
\beq \label{a3} \frac{\partial {\mathcal A}_3}{\partial t_1} +
({\bf v}_{\eta_3}\cdot \nabla_r){\mathcal A}_3 =
i\frac{\bar{\chi}_{\eta_1\eta_2\eta_3}({\bf q}_1,{\bf
q}_2)}{\omega_{\eta_3}} {\mathcal A}_1{\mathcal A}_2 \enq
where $\nabla_r = \frac{\partial}{\partial {\bf r}_1}$. The
nonlinear coefficient $\chi_{\eta_1\eta_2\eta_3}({\bf q}_1,{\bf
q}_2)$ is given by
\bee \label{chi1} \chi_{\eta_1\eta_2\eta_3}({\bf q}_1,{\bf q}_2) =
%\non\\
\sum_{{\bf n}_1,{\bf n}_2,{\bf n}_3}
\sum_{j_1,j_2,j_3}J_{3,j_1j_2j_3}({\bf n}_1,{\bf n}_2,{\bf
n}_3)\non\\     \times\bar{\phi}_{j_1,\eta_1}({\bf n}_1,{\bf
q}_1)\bar{\phi}_{j_2,\eta_2} ({\bf n}_2,{\bf
q}_2)\phi_{j_3,\eta_3}({\bf n}_3,{\bf q}_1 + {\bf q}_2) \ene
where $\phi_{j,\eta}({\bf n},{\bf q})$ are entries of the column
eigenvector $\Phi_\eta({\bf n},{\bf q})$, $j=1,2,...,\tilde{N}$,
(hereafter vectors of the reciprocal lattice ${\bf Q}$ are dropped
in the arguments  of the functions). In the particular case of
the second-harmonic generation, when conditions (\ref{second})
are satisfied one has
\beq \label{b1} \frac{\partial {\mathcal A}_1}{\partial t_1} +
({\bf v}_{\eta_1}\nabla_r){\mathcal A}_1 =
\frac{i\chi_{\eta_1\eta_2}}{\omega_{\eta_1}} \bar{{\mathcal
A}}_1{\mathcal A}_2,
% \enq
%
%\beq \label{b2}
\qquad
 \frac{\partial {\mathcal A}_2}{\partial t_1} +
({\bf v}_{\eta_2}\nabla_r){\mathcal A}_2 =
\frac{i\bar{\chi}_{\eta_1\eta_2}}{\omega_{\eta_2}} {\mathcal
A}_1^2 \enq
where for $\chi_{\eta_1\eta_2}\equiv \chi_{\eta_1\eta_2}({\bf q})$
one must use the formula
\beq \label{chi2} \chi_{\eta_1\eta_2} = \sum_{{\bf n}_1,{\bf
n}_2,{\bf n}_3} \sum_{j_1,j_2}J_{3,j_1j_1j_2}({\bf n}_1,{\bf
n}_2,{\bf n}_3)
%\times\non\\
{(\bar{\phi}_{j_1,\eta_1}({\bf n}_1,{\bf q}))}^2
\phi_{j_2,\eta_2}({\bf n}_2,2{\bf q}) \enq

\section{Second harmonic generation in a two-dimensional lattice.}

As the simplest example of the above approach we consider a 2D
diatomic lattice having atoms with masses ${M_1}$ and ${M_2}$
with linear interactions between nearest-neighbors and subject to
a nonlinear on-site potential.
%fig1%%%%%%%%%
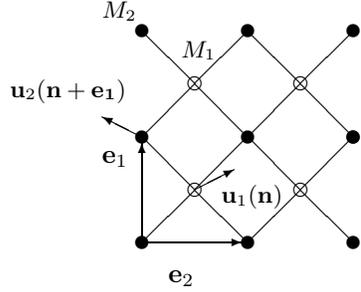
\begin{figure}[ht]
\label{f1}
\begin{picture}(10,40)
\put(90,-50){\vector(0,4){38}} \put(90,-50){\vector(4,0){38}}
\put(110,-30){\vector(2,1){15}} \put(90,-10){\vector(-2,1){15}}
\put(40,5){{\small${\bf u}_{2}(\bf{n}+\bf{e}_{1})$}}
\put(120,-35){{\small${\bf u}_{1}(\bf{n})$}}
\put(90,-50){\line(1,1){80}}
 \put(90,-10){\line(1,1){40}}
\put(130,-50){\line(1,1){40}} \put(130,-50){\line(-1,1){40}}
\put(170,-10){\line(-1,1){40}} %
\put(170,-50){\line(-1,1){80}} \put(90,-50){\circle*{5}}
\put(90,-10){\circle*{5}} \put(90,30){\circle*{5}}
 %\put(70,160){\circle*{5}}
\put(130,-50){\circle*{5}} \put(170,-50){\circle*{5}}
\put(170,-10){\circle*{5}}
%\put(190,40){\circle*{5}}
\put(110,-30){\circle{5}} \put(110,10){\circle{5}}
\put(130,-10){\circle*{5}} \put(130,30){\circle*{5}}
\put(150,-30){\circle{5}}
 \put(150,10){\circle{5}}
\put(170,-10){\circle*{5}} \put(170,30){\circle*{5}}
\put(100,-65){${\bf e}_{2}$} \put(75,-20){${\bf e}_{1}$}
\put(105,20){{\small $M_{1}$}} \put(75,35){{\small$ M_{2}$}}
\end{picture}
\vspace{3 true cm}
\caption{The example of a diatomic lattice. Solid lines
indicate inter-atomic interaction taken into account.}
\label{figone}
\end{figure}
In order to introduce atom numbering we can choose a cell as is
shown in Fig.~\ref{figone}. Then the lattice Hamiltonian reads
\bee H &= & \sum_{\bf n}\left\{\frac{{\bf p}^2_1({\bf n})}{2M_1} +
\frac{{\bf p}^2_2({\bf n})}{2M_2}
%\right.\non\\
+ \frac{K_{21}}{2}{\bf u}_1^2({\bf n})  + \frac{K_{22}}{2}{\bf
u}_2^2({\bf n})
    \right. \non\\
&+&\frac{K_2}{2}[({\bf u}_1({\bf n}) - {\bf u}_2({\bf n}))^2
    + ({\bf u}_1({\bf n}) - {\bf
u}_2({\bf n} + {\bf e}_1))^2
 \non\\
 &+& ({\bf u}_1({\bf n}) - {\bf
u}_2({\bf n} + {\bf e}_2))^2+ ({\bf u}_1({\bf n}) - {\bf u}_2({\bf
n} + {\bf e}_1 + {\bf e}_2))^2]
 \non\\
 &+& \left.\frac{K_{31}}{3}[u_{11}^3({\bf n}) +
u_{12}^3({\bf n})]
 + \frac{K_{32}}{3}[u_{21}^3({\bf n}) + u_{22}^3({\bf n})] \right\}
\ene
Here $K_{2}$ is a linear force constant, $K_{2j}$ and $K_{3j}$ are
respectively the linear and nonlinear constants of a force applied
to the j-th atom in a cell and originated by the on-site
potential, $u_{ij}$ is the $i$-th component of the displacement of
the $j$-th atom. We assume that the lattice is square with a
unitary lattice constant: ${\bf e}_i{\bf e}_j = \delta_{ij}$.
Then in order to represent the evolution equation in the form
(\ref{nlevol}) we define
%
%\beq
$
{\psi}({\bf n}) = \mbox{col}\,( \sqrt{M_1}u_{11},
\sqrt{M_2}u_{12}, \sqrt{M_1}u_{21}, \sqrt{M_2}u_{22} )
$
%\enq
%
and calculate the components of $J_2$ and $J_3$.  Then the
dispersion relation for the linear lattice can be written in the
form
\beq \label{displaw} \omega_{\eta}^2({\bf q}) =
\omega^2_{0,\sigma}\left[1 + (- 1)^{\kappa} \alpha_{\sigma}\sqrt{1
- \gamma_{\sigma}F(\bq)}\right] \enq
where $\kappa = \left[\frac{\eta+1}{2}\right]$, here $[\cdot]$
stands for the integer part, $\sigma = 2^{\frac{1 + (-
1)^{\eta}}{2}}$,
\beq F(\bq) = \sin^2(\frac{q_1}{2}) + \sin^2(\frac{q_2}{2}) +
\sin^2(\frac{q_1}{2})\sin^2(\frac{q_2}{2}), \enq
%
%
%\beq \alpha_{\sigma} = \sqrt{\frac{(1 - m)^2(k_{\sigma}^2 +
%2k_{\sigma}) + (1 + m)^2} {(1 + m)^2(1 + k_{\sigma})^2}}, \enq
%
\[
\gamma_{\sigma} = \frac{4m}{(1 - m)^2(k_{\sigma}^2 + 2k_{\sigma})
+ (1 + m)^2}, \qquad
\alpha_\sigma=\sqrt{\frac{4m}{\gamma_\sigma}}\frac{1}{(1+m)(1+k_\sigma)}
\]
$\omega^{2}_{0\sigma} = 2\omega_0^2(1 + k_{\sigma})(1 + m)$,
$k_{\sigma} = \frac{K_{2\sigma}}{4K_2}$, $m = \frac{M_2}{M_1}$ and
$\omega_0 = \sqrt{\frac{K_2}{M_2}}$, and  $\eta = 1,2$ correspond
to the low frequency branches which are reduced to the accoustic
modes at $K_{2j} = 0$, while $\eta=3,4$ denotes high frequency
optical branches.

Peculiarities of the second harmonic generation in the 2D case
follow from the dispersion relation. Firstly, in contrast to the
1D case the resonant conditions (\ref{second}) are satisfied on a
continuous line which can be seen from the fact that the equation
for the line can be written as $\sum_{ij}C_{ij}X^iY^j = 0$ with
$X = \sin^2(\frac{q_1}{2})$,  $Y = \sin^2(\frac{q_2}{2})$ and
constants $C_{ij}$ which are trivially found from (\ref{displaw}).
Secondly, energy transfer can occur between modes with the same
direction of velocities or with different directions of
velocities. In particular, the velocities can be orthogonal. The
former case in many aspects is analogous to the second harmonic
generation in the 1D chain (Konotop 1996, 1997). Finally, it is
possible to have {\em simultaneous} processes when one
low-frequency mode can resonantly interact with two other modes.
In particular the processes $2\omega_1({\bf q}_1) =
\omega_2(2{\bf q}_1)$ and $2\omega_1({\bf q}_1) = \omega_3(2{\bf
q}_1)$ can occur simultaneously for ${\bf q}_1 = (\pi,0)$ if the
lattice parameters are chosen as following $m = 12\frac{k_1^2 +
2k_1 + 1}{3k_1^2 + 6k_1 + 4}$ and $k_2 = \frac{12k_1^2 + 21k_1 +
10}{3(1+k_1)}$. Although in all other points the equations
(\ref{b1}) are similar to the equations describing the
second-harmonic generation in nonlinear optics, at such points
equations (\ref{b1}) are not valid any more.

The effective nonlinearity controlling SHG is now given by
\bee \label{chi3} \chi_{\eta_1\eta_2}({\bf q})  =
% \non \\
\frac{1}{\sqrt{M_{1}^3}}[K_{31}A^{*^{2}}_{\eta_{1},1}({\bf q})
A_{\eta_{2},1}(2{\bf q}) + K_{32}A^{*^{2}}_{\eta_{1},3}({\bf q})
A_{\eta_{2},3}(2{\bf q})] \non\\ +
\frac{1}{\sqrt{M_{2}^3}}[K_{31}A^{*^{2}}_{\eta_{1},2} ({\bf
q})A_{\eta_{2},2}(2{\bf q}) + K_{32}A^{*^{2}}_{\eta_{1},4} ({\bf
q})A_{\eta_{2},4}(2{\bf q})] \non \\ \ene
Here $A_{\eta,j}({\bf q})$ is the amplitude which can be trivially
found from (\ref{spectral2}).

If ${\bf v}_{\eta_1}$ is parallel to ${\bf v}_{\eta_2}$ then this
reduces the system (\ref{b1}) describing the evolution of the
envelope to an effectively one dimensional problem
\beq \label{b11} \frac{\partial{\mathcal A}_1}{\partial t_1} +
v_1\frac{\partial{\mathcal A}_1}{\partial x} =
i\frac{\chi}{\omega_1}\bar{{\mathcal A}}_1{\mathcal A}_2,
% \enq
%
%\beq \label{b21}
\qquad \frac{\partial{\mathcal A}_2}{\partial t_1} +
v_2\frac{\partial{\mathcal A}_2}{\partial x} =
i\frac{\bar{\chi}}{\omega_2}{\mathcal A}^2_1 \enq
where $x$ is the slow spatial coordinate in the direction of the
group velocity vector and $\chi=\chi_{\eta_1\eta_2}({\bf q})$.
This system is well studied in the context of nonlinear optics
(see e.g. Suhorukov. 1988). It also reduces the 2D problem to
quasi-one-dimensional one (Konotop, 1996).

\subsection{Case of orthogonal velocities}

If the velocities of the first and second harmonics are
orthogonal then system (\ref{b1}) is reduced to
\beq \label{b12} \frac{\partial{\mathcal A}_1}{\partial t_1} +
v_1\frac{\partial {\mathcal A}_1}{\partial  x} =
i\frac{\chi}{\omega_1} \bar{{\mathcal A}}_1{\mathcal A}_2, \qquad
% \enq
%
%\beq \label{b22}
\frac{\partial{\mathcal A}_2}{\partial t_1} + v_2\frac{\partial
{\mathcal A}_2}{\partial y} = i\frac{\bar{\chi}}{\omega_2}
{\mathcal A}^2_1 \enq
where $x$ and $y$ are spatial coordinates in the direction of the
group velocities of the first and second harmonic, respectively.

In a special case the system (\ref{b12}) can be linearized, using
the ideas of (Suhorukov. 1988). Indeed, let us introduce
normalized amplitudes
\[
a_1=\frac{\bar{\chi}}{\sqrt{2}\omega_1}e^{-i\alpha_1} {\mathcal
A}_1,\qquad a_2=\frac{\bar{\chi}}{\omega_1}e^{-i\alpha_2}{\mathcal
A}_2, \qquad 2\alpha_1-\alpha_2=\pi/2
\]
and look for a solution of (\ref{b12}) such that the functions
$a_1$ and $a_2$ are real. Let us also define operators
\[
\hat{L}_1=\frac{\partial}{\partial t} +v_1\frac{\partial}{\partial
x}, \qquad \hat{L}_2=\frac{\partial}{\partial t}
+v_2\frac{\partial}{\partial y}
\]
Then system (\ref{b12}) can be rewritten in the form
\begin{equation}
\label{liner1} \hat{L}_1a_1=a_1a_2,\qquad \hat{L}_2a_2=-a_1^2.
\end{equation}
Next we apply $\hat{L}_1$ to the second of the above equations. As
a result we arrive at $\hat{L}_2(\hat{L}_1a_2- a_2^2)=0$. This
means that
%\begin{equation}
%\label{liner2}
$ \hat{L}_1a_2- a_2^2=C(x,y-v_2t) $
%\end{equation}
where $C(\cdot,\cdot)$ is a function which is determined from the
boundary (initial) conditions. Then the linear equation for
$1/a_1$ takes the form
\begin{equation}
\label{liner3} \hat{L}_1^2\frac{1}{a_1} +
C(x,y-v_2t)\frac{1}{a_1}=0.
\end{equation}

\subsubsection{Solitary wave solution}

The first simplification of (\ref{liner3}) can be achieved by
assuming that $C(\cdot,\cdot)$ does not depend on $x$:
$C(\cdot,\cdot)\equiv C(\xi_2)$. Here we introduce running
variables: $\xi_1=x/v_1-t$ and $\xi_2=y/v_2-t$. Then one can look
for the solutions depending on the running variables only:
$a_j\equiv a_j(\xi_1,\xi_2)$. Here we represent one of them which
is generated by the Lorentz pulse
$C(\xi_2)=$const$/(1+\xi_2^2/v_2^2)^2$. It reads
\begin{eqnarray}
\non {\mathcal A}_1&=&\frac{\sqrt{2}\omega_1}{\bar{\chi}
\sqrt{1+\xi_2^2}[ \sinh \xi_1 \sinh(\arctan \xi_2)-\cosh \xi_1
\cosh(\arctan \xi_2) ]}
\\
\non {\mathcal
A}_2&=&\frac{i\omega_1}{\bar{\chi}}\frac{\xi_2\cosh(\xi_1-\arctan
\xi_2)-\sinh(\xi_1-\arctan \xi_2)}{(1+\xi_2^2)\cosh(\xi_1-\arctan
\xi_2)}
\end{eqnarray}

The first harmonic represents a solitary wave pulse which is
exponentially localized in the direction of the group velocity of
the first mode and displays power-law-localization in the
direction of propagation of the second harmonic. The second
harmonic is delocalized in the direction of the group velocity of
the first mode. In that direction it has a kink-like shape.  At
$t>0$ the two modes being coupled with each other move without
distortion  in the direction defined by $x/v_1+y/v_2=0$.

\subsubsection{Spatial structure}

In order to illustrate the possibility of existence of stationary
spatial structures we consider solutions of (\ref{b12}) [or
(\ref{liner1})] in a space defined by $x\geq 0$, $y\geq 0$ when
$\frac {\partial {\mathcal A}_j}{\partial t_1} = 0$ subject to
proper boundary conditions  and subject to the assumption that $
v_1v_2>0$. It reads ($x_0$ and $y_0$ are real constants)
   It reads
\begin{eqnarray}
\non  {\mathcal A}_1(x,y) &=& \frac{2\omega_1\sqrt{v_1v_2}}
{\bar{\chi}
\left(e^{x+x_0}\sinh(y+y_0)+e^{-x-x_0}\cosh(y+y_0)\right)}
 \\ \non {\mathcal A}_2(x,y) &=&
\frac{v_1\omega_1\left(e^{-x-x_0}\cosh(y+y_0)-e^{x+x_0}\sinh(y+y_0)\right)}
{\bar{\chi}\left(e^{x+x_0}\sinh(y+y_0)+e^{-x-x_0}\cosh(y+y_0)\right)}
 \end{eqnarray}
The respective solution represents energy transfer form the first
harmonic, localized in the direction of the group velocity of the
first mode, to the second harmonic, in the same direction having
a kink-like shape, which occurs in the direction of the group
velocity of the second mode.

\section{Quadratic solitons in a lattice}

Consideration provided in the previous section does not take into
account the dispersion which is intrinsic property of the
nonlinear lattices. This was justified by the choice of the
scaling of the problem. Let us now turn to the case when the group
velocity dispersion appears to be an important factor (Konotop
and Malomed, 2000). Namely, assuming that resonant condition
(\ref{second}) is satisfied let us look for the solution of
nonlinear model (\ref{nlevol}) in the form of the expansion [c.f.
(\ref{expan})]
\begin{equation}
\label{expan1}
 \Psi({\bf n}) =\epsilon^2 \Psi^{(0)}({\bf n}_0,
\{{\bf r}_{\nu}\};\{t_{\nu}\}) + \epsilon^3 \Psi^{(1)}({\bf n}_0,
\{{\bf r}_{\nu}\};\{t_{\nu}\}) + \cdots
\end{equation}
Then the substitution
\begin{equation}
\label{subs1}
  \psi^{(0)}({\bf n}_0, \{{\bf
r}_{\nu}\};\{t_{\nu}\}) = \sum_{\alpha = 1}^2{\mathcal
A}_{\alpha}({\bf r}_{1};t_{1}) \phi_{\eta_\alpha}({\bf n}_0;{\bf
q}_\alpha)e^{-i\omega_{\eta_\alpha}({\bf q}_\alpha)t_0} + c.c.
\end{equation}
results in the  evolution equations for the amplitudes (they are
obtained in the third order of the asymptotic expansion)
\begin{equation}
\label{qs1} i\frac{\partial A_{\eta_1}}{\partial t_2}+\frac 12
\sum_{i,j}\omega_{\eta_1}^{ij}\frac{\partial^2
A_{\eta_1}}{\partial x^i\partial
x^j}+\chi_{\eta_1}\bar{A}_{\eta_1} A_{\eta_2}=0
\end{equation}
\begin{equation}
\label{qs2} i\frac{\partial A_{\eta_2}}{\partial t_2}+\frac 12
\sum_{i,j}\omega_{\eta_2}^{ij}\frac{\partial^2
A_{\eta_2}}{\partial x^i\partial x^j}+\chi_{\eta_2}A_{\eta_1}^2=0
\end{equation}
Some particular solutions of these equations (however in
applications to the nonlinear optics) have been considered in
(Malomed et al. 1997).

\section{Dynamics of a double chains}

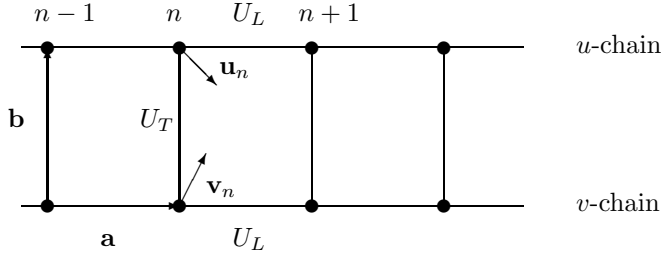
\begin{figure}[h]
 \begin{picture}(200,100)
\put(60,20){\vector(1,2){10}} \put(60,80){\vector(1,-1){14}}
%\put(90,60){\vector(2,1){15}} \put(70,80){\vector(-2,1){15}}
%\put(20,95){{\small${\bf u}_{2}(\bf{n}+\bf{e}_{1})$}}
%\put(100,55){{\small${\bf u}_{1}(\bf{n})$}}
%
\put(0,20){\line(1,0){190}} \put(0,80){\line(1,0){190}}
\put(10,20){\vector(1,0){50}} \put(10,20){\vector(0,1){60}}
\put(60,20){\line(0,1){60}} \put(110,20){\line(0,1){60}}
\put(160,20){\line(0,1){60}} \put(10,20){\circle*{5}}
\put(60,20){\circle*{5}} \put(110,20){\circle*{5}}
\put(160,20){\circle*{5}} \put(10,80){\circle*{5}}
\put(60,80){\circle*{5}} \put(110,80){\circle*{5}}
\put(160,80){\circle*{5}}
\put(70,25){${\bf v}_{n}$} \put(75,70){${\bf u}_{n}$}
\put(80,5){$U_{L}$} \put(80,90){$U_{L}$} \put(45,50){$U_{T}$}
\put(30,5){${\bf a}$} \put(-5,50){${\bf b}$} \put(5,90){$ n-1$}
\put(55,90){$n$} \put(105,90){$n+1$} \put(210,18){$v$-chain}
\put(210,78){$u$-chain}
\end{picture}
 %\vspace{5mm}
\caption{Schematic representation of a double chain.
  The axes $x$ and $y$ will be chosen along ${\bf
a}$ and ${\bf b}$, respectively.} \label{figtwo}
\end{figure}

In the present section we deal with phenomena originated by
resonant interactions of longitudinal and transverse degrees of
freedom in a  system consisting of two coupled vector atomic chain
depicted in Fig.~\ref{figtwo}.
 %fig1%%%%%%%%%

The potential energy of the interaction between the neighbor
sites, say $n$ and $n+1$, is considered to be $U_L(|{\bf a}+{\bf
u}_{n+1}-{\bf u}_n|)$ in the $u$-chain  and $U_L(|{\bf a}+{\bf
v}_{n+1}-{\bf v}_n|)$ in the $v$-chain. The nearest neighbors in
the both chains interact with the energy $U_T(|{\bf b}+{\bf
u}_n-{\bf v}_n|)$.

Thus the hamiltonian of the system can be written down in the form
\begin{eqnarray}
\label{hamilt} H&=&\sum_n\left\{\frac{{\bf p}_{u,n}^2}{2m}+
\frac{{\bf p}_{v,n}^2}{2m}+U_L(|{\bf a}+{\bf u}_{n+1}-{\bf
u}_n|)  \right.\non \\
  &+&\left. U_L(|{\bf a}+{\bf v}_{n+1}-{\bf v}_n|)+ U_T(|{\bf b}+{\bf u}_n-{\bf v}_n|)\right\}
\end{eqnarray}
The vectors ${\bf u}_n=(u_{x,n},u_{y,n})$ and ${\bf
v}_n=(v_{x,n},v_{y,n})$ are displacements of the atoms from their
equilibrium positions in the  $u$- and $v$-lattices, respectively.
Without restriction of generality, in what follows the atomic mass
is considered to be one: $m=1$.

If $\omega_u$ and $\omega_v$ are a characteristic frequencies of
excitations propagating along u- and v-chains.  and $\omega_t$ is
a frequency of the transverse oscillations, it is natural to
expect enhancement of the interaction when
$n_t\omega_t+n_u\omega_u+n_v\omega_v=0$, $n_t$, $n_u$, and $n_v$,
being integers. The "lowest" resonant phenomena will be related
to resonant three wave interactions either $\omega_t=2\omega_l$
or $\omega_l=2\omega_t$.

We will concentrate on excitation of a relatively small amplitude.
Then the potential energy in (\ref{hamilt}) can be expanded in the
Taylor series
\begin{eqnarray}
\non  U_L(|{\bf a}+{\bf u}_{n+1}-{\bf u}_n|)&=&\frac 12
K_2(L_u(n)-a)^2+\frac 16 K_3(L_u(n)-a)^3+ ...
\\ \non
 U_T(|{\bf b}+{\bf u}_n-{\bf v}_n|)&=&\frac 12 Q_2(L(n)-b)^2+\frac
16 Q_3(L(n)-b)^3+ ...
\end{eqnarray}
where
\[
 L_u(n)=\left[(a+u_{x,n}-u_{x,n-1})^2+(u_{y,n}-u_{y,n-1})^2
\right]^{1/2}
\]
 is the distance between two neighbor sites in the
u-chain ($L_v$ is defined by analogy)
\[
L(n)=\left[(u_{x,n}-v_{x,n})^2+(b+u_{y,n}-v_{y,n})^2 \right]^{1/2}
\]
is the distance between neighbor sites u- and v-lattices  and the
force coefficients are given by $K_n\equiv
\frac{d^nU_L(a)}{da^n}$, and $Q_n\equiv\frac{d^nU_T(b)}{db^n}$
($K_2$ and $Q_2$ being positive).

Since we are dealing with the phenomena originated by resonant
three wave interactions the terms due to the nonlinearity of the
central-force potentials $U_L$ and $U_T$ are not important in the
leading order, since the lowest order nonlinearity results from
the expansion of $L_u(n)$, $L_v(n)$ and $L(n)$ with respect to
small displacements of atoms from the equilibrium positions. Then
the equations of motion read
\begin{eqnarray}
\label{evol_ux} \ddot{u}_{x,n}&=&K_2(u_{x,n+1}+u_{x,n-1}-2u_{x,n})
  - \frac{Q_2}{b}(u_{y,n}-v_{y,n})(u_{x,n}-v_{x,n})\nonumber
\\ &+&\frac{K_2}{2a}[(u_{y,n+1}-u_{y,n})^2-(u_{y,n-1}-u_{y,n})^2]
\end{eqnarray}
\begin{eqnarray}
\label{evol_vx} \ddot{v}_{x,n}&=&K_2(v_{x,n+1}+v_{x,n-1}-2v_{x,n})
- \frac{Q_2}{b} (v_{y,n}-u_{y,n})(v_{x,n}-u_{x,n}) \nonumber
\\ &+&
\frac{K_2}{2a}[(v_{y,n+1}-v_{y,n})^2-(v_{y,n-1}-v_{y,n})^2]
\end{eqnarray}
\begin{eqnarray}
\label{evol_uv}
\ddot{u}_{y,n}=-Q_2(u_{y,n}-v_{y,n})-\frac{Q_2}{2b}
(u_{x,n}-v_{x,n})^2 +  \nonumber \\    \frac{K_2}{a}
\left[(u_{x,n+1}-u_{x,n})(u_{y,n+1}-u_{y,n})
-(u_{x,n}-u_{x,n-1})(u_{y,n}-u_{y,n-1})\right] \nonumber \\
\end{eqnarray}
\begin{eqnarray}
\label{evol_vu}
\ddot{v}_{y,n}=-Q_2(v_{y,n}-u_{y,n})+\frac{Q_2}{2b}
(u_{x,n}-v_{x,n})^2 - \nonumber
\\ \frac{K_2}{a}\left[(v_{x,n}-v_{x,n-1})(v_{y,n}-v_{y,n-1})+
(v_{x,n+1}-v_{x,n})(v_{y,n+1}-v_{y,n})\right] \nonumber \\
\end{eqnarray}

\subsection{Resonance $\omega_t=2\omega_l$.}

As an example consider the situation when motion of u- and v-
chains is synchronized, i.e. $\omega_u=\omega_v=\omega$  and
respectively $\omega_t=2\omega_l$. As far as lattices are
considered identical the above assumption means that the wave
vectors of the carrier wave (cw) of excitations propagating along
both chains coincide. We will concentrate on the case when the
wave vector of a single lattice borders the boundary of the
Brillouin zone. Then solutions of the hamiltonian equations for
the lattice displacements can be represented in the form
\begin{eqnarray*}
%\label{expansion}
& & u_{x,n}(t)=\epsilon^2 \left[u_x^{(1)}(t_1;x_1)
 (-1)^ne^{-i\omega t_0}+O(\epsilon^2)\right]+ c.c.
\\
& &v_{x,n}(t)=\epsilon^2 \left[v_x^{(1)}(t_1;x_1)
(-1)^ne^{-i\omega t_0}+ O(\epsilon^2)\right]+ c.c.
\\
& &u_{y,n}(t)=\epsilon^2 \left[w^{(1)}(t_1;x_1)e^{-2i\omega t_0} +
O(\epsilon^2)\right]
 + c.c. \\
& &v_{y,n}(t)=-\epsilon^2\left[w^{(1)}(t_1;x_1) e^{-2i\omega t_0}+
O(\epsilon^2)\right]
 + c.c.
\end{eqnarray*}
Here we introduce a hierarchy of slow times $t_j=\epsilon^j t$ and
spatial scales $n_j=\epsilon^j n$ ($j=0,1,...$). The scales $n_j$
($j\geq 1$) are slow ones and that is why we introduce $x_j=an_j$
($j\geq 1$) which are regarded as continuous variables. The symbol
$O(\epsilon^2)$ stands for all the terms of order of $\epsilon^2$
which have frequencies either different from $\omega$ for
longitudinal components or different from $2\omega$ for transverse
component. Also in the expansion we use the convention that in the
arguments of the functions only the most "rapid" scales are
indicated [so, for example $u_1(t_1;x_1)\equiv
u_1(t_1,t_2,...;x_1,x_2,...)$]. "c.c." stands for the complex
conjugation. i.e. Quantity $2w(\xi;t_2,x_2)$ can be interpreted as
an is an enlargement of the distance between neighbor atoms in
different chains which oscillate with opposite phases.

  In order to get solitonic effects this expansion should be
provided up to the terms of order of $\epsilon^5$. After that the
expansion for ${\bf u}$, ${\bf v}$ must be substituted in the
Hamiltonian equations and terms of the same order of the small
parameter must be collected. Dropping this straightforward
standard calculus we pass to the description of the equations
governing the dynamics of the double chain.

In the lowest order with respect to the small parameter (i.e. in
the $\epsilon^2$ order)  the chains appear to be independent. The
respective terms of the expansion of the longitudinal and
transverse components give the two values for the frequency $
 \omega= 2\sqrt{K_2}$ and $\omega=\sqrt{2Q_2}$,
which must coincide. In other words the phenomenon can be observed
subject to the condition $
 \frac{d^2U_T(b)}{db^2}=2\frac{d^2U_L(a)}{da^2}.
$ Subject to the above conditions, the equations of the third
order with respect to epsilon become zero identically. Thus,
computing the terms of order of $\epsilon^4$ one arrives at the
system of equations
\begin{equation}
\label{u} -i\frac{\partial U}{\partial \tau} +\frac{\partial^2 U}
{\partial x_1^2}+ 2(\bar{U}-\bar{V})W=0,
%\end{equation}
%\begin{equation}
%\label{v}
\qquad
 -i\frac{\partial V}{\partial \tau} +\frac{\partial^2 V}
{\partial x_1^2}+ 2(\bar{V}-\bar{U})W=0
\end{equation}
\begin{equation}
\label{w} -i\frac{\partial W}{\partial \tau} +\frac{1}{2}(U-V)^2=0
\end{equation}
where $U(\tau, x_1)=\frac{4}{b}u_x^{(1)}(t_2,x_1)$,  $V(\tau, x_1
)=\frac{4}{b}v_x^{(1)}(t_2,x_1)$, $W(\tau,
x_1)=\frac{8}{b}w(t_2,x_1)$ and $\tau=\sqrt{K_2}t_2/2$.

Let us consider a particular solution of the above system, which
corresponds to out-of-phase oscillations of the atoms in the u-
and v-chains: $U=-V$. Then the system (\ref{u})-(\ref{w}) takes
the form
\begin{equation}
\label{u1} -i\frac{\partial U}{\partial \tau} +\frac{\partial^2 U}
{\partial x_1^2}+ 4\bar{U}W=0, \qquad
  -i\frac{\partial W}{\partial \tau} +2U^2=0
\end{equation}
We notice that the hamiltonian governing this system is as follows
\begin{equation}
\label{hamilt1} H=\int
\left(|U_x|^2+2\overline{U}^2W+2\overline{W}U^2\right)\,d x
\end{equation}

System (\ref{u1}) admits  solutions in a form of two coupled dark
solitons
\begin{equation}
\label{dark1} U=\beta^2e^{2i\beta^2 t}\tanh(\beta x),\qquad
W=-\frac{\beta^2}{2}e^{4i\beta^2 t}\tanh^2(\beta x)
\end{equation}
and two coupled bright solitons
\begin{equation}
\label{bright1} U=\frac{\beta^2}{\sqrt{2}}\frac{e^{-i\beta^2
t}}{\cosh(\beta x)},\qquad W=-\frac{\beta^2}{2}\frac{e^{-2i\beta^2
t}}{\cosh^2(\beta x)}
\end{equation}
These solutions describe localized in space narrowing and
broadening of the double chain in transverse direction.  The
respective trajectories of the atoms are parabolas which are
modulate by the envelopes (\ref{dark1}) and (\ref{bright1})
respectively.

\section{Conclusion}

In the present paper we have considered several effects in
multidimensional lattices which are originated by resonant wave
interactions.  All the examples were given for nonlinear
Klein-Gordon like lattices. The theory however allows for
straightforward modification for the sake of description of
lattices of the nonlinear Schr\"{o}dinger type (see e.g. Konotop
1997; Konotop, {\it et al.} 1999). Also, resonant muti-wave
interactions in lattice can be considered in the similar way.

Other relevant effects we did not discussed here are associated
with the phase mismatch (i.e. when, for example,
(\ref{threewave}) is substituted by
$\omega_3=\omega_1+\omega_2+\Delta\omega,$
$\bq_3=\bq_1+\bq_2+{\bf Q}$, where $\Delta\omega\ll \omega_j$
$(j=1,2,3)$.

Finally, taking into account that resonant mode interactions is a
traditional subject of the nonlinear optics it is worth pointing
out that the difference between lattices and optical media comes
from the anisotropy, which is an intrinsic property of the
lattice. Also, lattices provide physical systems where almost all
relations among parameters are available (including change of the
signs of nonlinear and dispersive terms).

\section*{Acknowledgments}
Author is indebt to B. A. Malomed and to A. Gon\c{c}alves for
collaboration. Support from FEDER and Program PRAXIS XXI, grant
No Praxis/P/Fis/10279/1998 is acknowledged.

\end{document}